\documentclass[letterpaper, 10 pt, conference]{ieeeconf}  %

\IEEEoverridecommandlockouts                              %
\usepackage{amsmath,amssymb,amsfonts,amsthm}
\usepackage{bm,bbm,siunitx,hyperref,graphicx}
\usepackage{grffile}
\usepackage{float}	%
\usepackage{xcolor}
\usepackage{mathtools, nccmath}

\let\labelindent\relax%

\usepackage{enumitem}
\usepackage{makecell}
\usepackage{cellspace}
\usepackage{cite}
\usepackage{balance}   %
\usepackage[font=small,skip=20pt]{caption}	%

\usepackage{makecell}

\usepackage{multirow}

\usepackage{soul}	%

\usepackage{subcaption}

\usepackage{amsthm}	%

\DeclareSIUnit \ampereHour {Ah}
\DeclareSIUnit \wattHour {Wh}

\DeclareSIPostPower\powerThreeHalfs{\frac{3}{2}}

\theoremstyle{plain}

\theoremstyle{remark}

\theoremstyle{remark}

\theoremstyle{remark}

\theoremstyle{remark}

\newcommand{\mrb}[1]{\left( #1 \right)} %
\newcommand{\msb}[1]{\left[ #1 \right]} %
\newcommand{\mcb}[1]{\left\{ #1 \right\}} %

\newcommand{\figref}[1]{Fig.~\ref{#1}}
\newcommand{\secref}[1]{Section~\ref{#1}}
\newcommand{\tabref}[1]{Table~\ref{#1}}

\newcommand{\RealCoords}{\mathbb{R}^3}

\newcommand{\tetherForce}{\mathbf{t}}

\newcommand{\droneMass}{m}

\newcommand{\powerConst}{c_p}
\newcommand{\droneThrust}{f}

\newcommand{\tetherMassPerLength}{\lambda}
\newcommand{\tetherResistancePerLength}{\rho}

\newcommand{\tetherResistancePerLengthj}{\rho_j}
\newcommand{\tetherLength}{L}

\newcommand{\tetherLengthj}{l_j}
\newcommand{\tetherResistance}{R}

\newcommand{\tetherResistancej}{R_j}
\newcommand{\tetherResistancek}{R_k}

\newcommand{\droneVoltage}{V}
\newcommand{\droneVoltagej}{V_j}
\newcommand{\droneVoltagek}{V_k}

\newcommand{\droneCurrent}{i}
\newcommand{\droneCurrentj}{i_j}

\newcommand{\droneCurrentl}{i_l}
\newcommand{\dronePoweri}{P_i}
\newcommand{\dronePowerj}{P_j}

\newcommand{\sourceVoltage}{V_s}
\newcommand{\sourceCurrent}{i_s}
\newcommand{\sourcePower}{P_s}

\newcommand{\dronePower}{P}

\newcommand{\propSize}{A_\mathrm{prop}}
\newcommand{\airDensity}{\rho_\mathrm{air}}
\newcommand{\propEfficiency}{\eta_\mathrm{prop}}
\newcommand{\powertrainEfficiency}{\eta_\mathrm{pt}}
\newcommand{\nProps}{n}

\newcommand{\COMMENTOUT}[1]{}

\title{\LARGE \bf
Tethered Power for a Series of Quadcopters: Analysis and Applications}

\author{
	Karan P. Jain$^{*}$, Prasanth Kotaru$^{*}$, Massimiliano de Sa, Koushil Sreenath, and Mark W. Mueller%
	\thanks{
		\hspace{-8.5pt}$^*$denotes equal contribution and listed alphabetically. \newline
		The authors are with the Dept. of Mechanical Engineering, UC Berkeley, CA 94720, USA.
		{\tt\small \{karanjain, prasanth.kotaru, mz.desa, mwm, koushils\}@berkeley.edu}
	}
}%

\begin{document}
\setlength{\abovedisplayskip}{1pt}
\setlength{\belowdisplayskip}{0pt}%
\setlength{\textfloatsep}{1pt}	
\setlength{\abovecaptionskip}{1pt}

\maketitle

\begin{abstract}
Tethered quadcopters are used for extended flight operations where the power to the system is provided via a tether connected to an external power source.
In this work, we consider a system of multiple quadcopters powered by a single tether.
We study the design factors that influence the power requirements, such as the electrical resistance of the tether, input voltage, and quadcopters' positions.
We present an analysis to predict the required power to be supplied to a series of $N$ tethered quadcopters, with respect to the thrust of each quadcopter which guarantees electrical safety and helps in design optimization.
We find that there is a critical boundary of thrusts that cannot be exceeded due to fundamental electrical limitations.
We compare the power consumption for one tethered quadcopter and two tethered quadcopters and show that for large quadcopters far enough from the anchor point, a two-quadcopter system consumes lesser power.
We show that, for a representative use case of firefighting, a tethered system with two quadcopters consumes $26\%$ less power than a corresponding system with one quadcopter.
Finally, we present experiments demonstrating the use of a two-quadcopter tethered system as compared to a one-quadcopter tethered system in a cluttered environment, such as passing through a window and grasping an object over an obstacle.
\end{abstract}

\section{Introduction}\label{sec:intro}
Aerial vehicles like quadcopters are used in various applications, from surveillance to manipulation to exploring other planets\cite{grip2019flight, lorenz2018dragonfly}. While these vehicles are primarily utilized for passive tasks such as surveillance\cite{jaimes2008approach} and photography\cite{cheng2015aerial}, research groups and industries actively pursue aerial vehicles for manipulation tasks involving grasping/positioning\cite{fishman2021dynamic}, assembling/dismantling parts\cite{augugliaro2014flight}, or transporting payloads\cite{ICC2018_MultiQuadLoad_FlexCable} using one or more vehicles. 

Aerial vehicles, however, are largely constrained in their flight time, and payload capacity, due to their limited hardware and power supply. 
Various innovations towards extending the flight time and range of quadcopters have been explored.
Methods such as swapping the batteries at a ground station\cite{lee2015autonomous}, replacing batteries in-flight using other quadcopters\cite{jain2020flying}, and using the batteries in multiple stages\cite{jain2020staging} have demonstrated increased flight times.
However, these systems can run out of power if not replaced in time and would require the quadcopters to land.

An alternate approach to extend flight time is to supply the power through an external tether from a fixed/mobile ground station. 
While a tethered quadcopter has limited flight reach and maneuverability, in applications such as atmospheric analysis\cite{rico2021trajectory}, construction/industrial inspections\cite{watanabe2016development}, surveillance\cite{torgesen2021airborne}, or aerial manipulation, the choice of tethering the quadcopter is a reasonable trade-off between reach and extended flight.
Their uses have also been demonstrated in a variety of commercial applications such as picking fruits in an orchard \cite{maor2021device} and cleaning buildings and wind turbine blades. 

Most tethered aerial systems are limited to only vertical flights, especially when carrying heavy tethers such as in cleaning buildings. 
A single tethered system is limited 
in their horizontal reachability. Works such as \cite{kosarnovsky2019string, bolognini2020lidar} consider a new type of system consisting of a series of tethered quadcopters to improve the reachability of the tethered systems, especially in cluttered environments. 

\begin{figure}[t]
	\centering
	\includegraphics[width=\columnwidth]{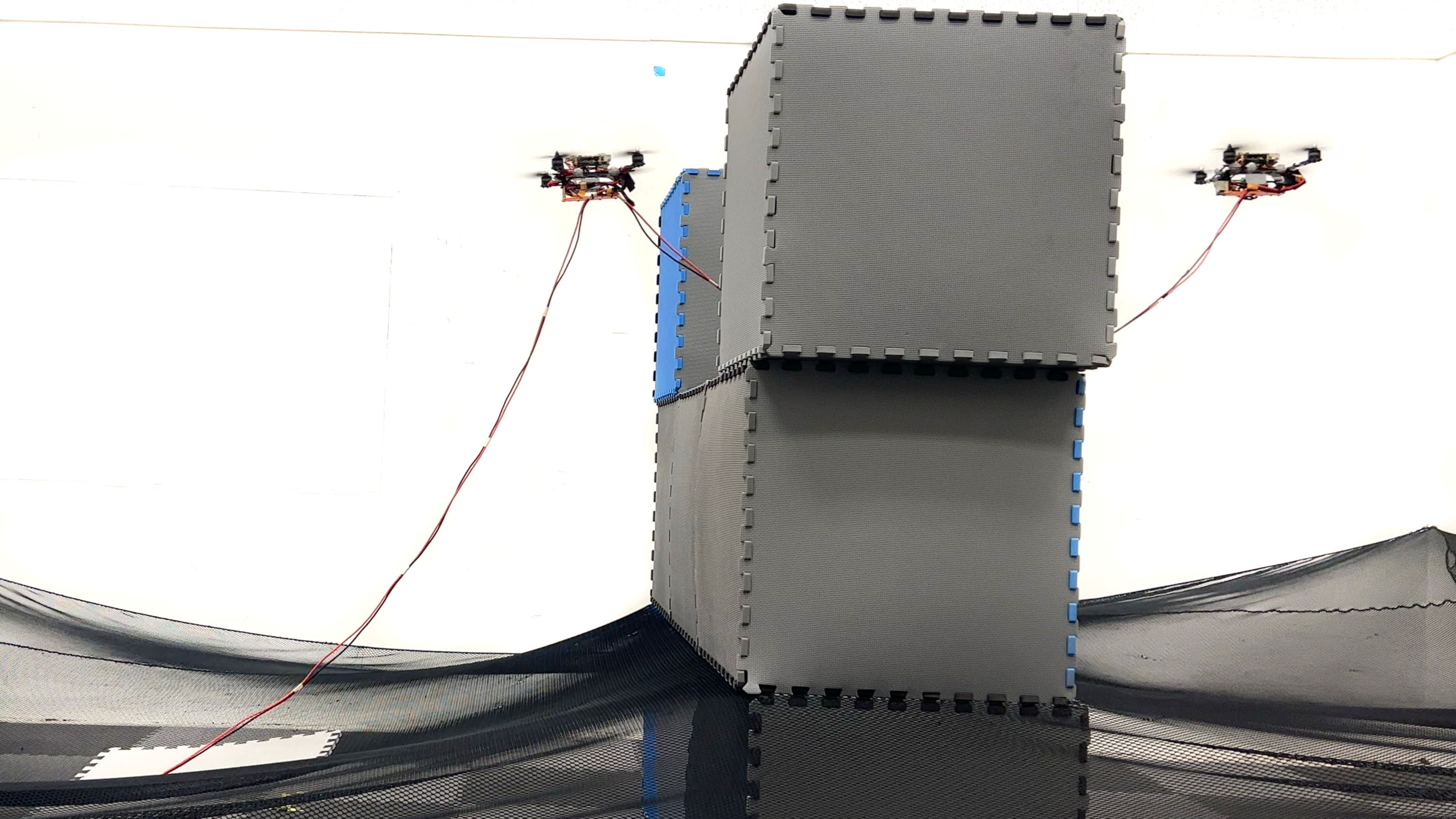}
	\caption{Experiment demonstrating two quadcopters tethered using a single cable, supplying power from an external power source. The quadcopters are electrically connected in parallel. Multiple quadcopters can be used collaboratively to achieve tethered flight over unknown/challenging terrain while increasing the horizontal reachability of the quadcopter.
		A video for the experiment is provided as an attachment.
	}
	\label{fig:introduction}
\end{figure}

\subsection{Related Work}
Quadcopter systems tethered to a fixed or moving base have been explored in the past \cite{walendziuk2020power, zikou2015power, tognon2016observer, tognon2020theory}. Control algorithms for the tethered quadcopters were developed in \cite{nicotra2014taut}, where the tether is modeled as a massless rigid link. 
To account for the mass of the cable during dynamic maneuvers, the tether is modeled as a series of lumped mass links in \cite{lee2015geometric, kotaru2018differential}. Catenary models are used to model the tether in \cite{d2021catenary, abhishek2021towards}, however, such models can be used only under quasi-static conditions. A system of tethered quadcopters connected in series is studied in \cite{kosarnovsky2019string, bolognini2020lidar, kotaru2020multiple}. A string of quadcopters would extend the horizontal reachability as well as the ability to navigate in a cluttered environment. 

Although tethering a quadcopter to a power supply provides access to unlimited energy, a power analysis is useful to guarantee that the power draw of the system is within the rated limit of the supply.
It is also beneficial in choosing the right type of tether since there is a trade-off between the mass of the tether (influencing the thrust and power requirements on the quadcopter) and the resistance of the tether (influencing the resistive power losses).

Energy analysis for a downward tethered quadcopter that is not externally powered is shown in \cite{lee2021energy}.
Power supply for tethered drones is analyzed in \cite{walendziuk2020power}, where comparisons between a battery-operated drone and a tethered drone with an external DC power source are drawn.
In, \cite{kiribayashi2015modeling}, power analysis for a single tethered quadcopter with respect to various factors such as input voltage, wire resistance, and cable length has been discussed, but the effects of catenary forces on the power were not considered. Moreover, extending the power to multiple quadcopters increases the complexity of the system and requires further study.

\subsection{Contributions}
In this work, we look into multiple quadcopters powered via a single tether from an external power source. The contributions of this work are as follows:
\begin{itemize}
	\item We formulate the power consumption as a function of quadcopter thrusts and study the influence of various parameters such as tether resistance and input voltage. The power analysis shows the existence of a critical boundary of thrusts beyond which it is electrically impossible to produce the thrusts.
	\item We show a power comparison for a fixed end-effector quadcopter position (see \figref{fig:dynamicsSchematicMultiDrone}) between one tethered quadcopter and a two tethered quadcopter system, taking into account the power analysis and catenary forces from the tether. This allows us to find configurations where two quadcopter system has a lower total power consumption that frees up thrust capacity for the end-effector quadcopter, increasing its maneuverability and agility.
	\item Finally, we demonstrate applications of a system of two tethered quadcopters -- flying through a corridor and picking up an object in a space with obstacles.
	\item Open source code for the power analysis is included.\footnote{after the manuscript gets accepted}
\end{itemize}

\section{System Description}
\label{sec:description}
\begin{figure}
	\centering
	\includegraphics[width=\columnwidth]{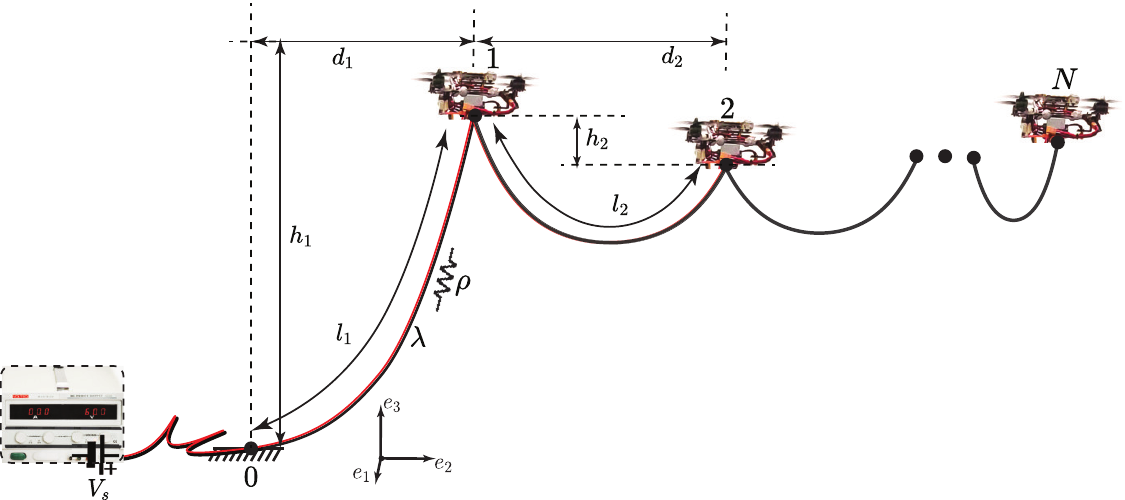}
	\caption{
		Schematic for a series of quadcopters in hover tethered to an external power source using a single tether.
		The last ($N$-th) quadcopter is referred to as the end-effector quadcopter.
	}
	\label{fig:dynamicsSchematicMultiDrone}
\end{figure}

In this section, we describe a series of identical quadcopters powered by a single tether connected to a power source and review various factors influencing the power requirement of the system under quasi-static hover conditions.

Consider a uniform tether supply connected to an external power source attached to the quadcopters as shown in \figref{fig:dynamicsSchematicMultiDrone}. 
One end of the tether is attached to a quadcopter and the other is fixed to the ground (indexed 0). We assume the tether is rigidly attached to the center-of-mass of the quadcopter, i.e., the tether applies only translation force on the quadcopter. %
Each quadcopter is indexed $i=\{1,2,\hdots,N\}$, in ascending order away from the power source.

Let the length of the cable segment between the quadcopters $i{-}1$ and $i$ be $l_i$. The horizontal and vertical distances between the quadcopters are given as $h_i$ and $d_i$, respectively. Without loss of generality, we consider the system to be in a single catenary plane for the rest of the paper. The total length of the tether from the power source to the last quadcopter $N$ is $L$. 
We assume the tether is uniform along its length, with $\tetherMassPerLength$ mass per unit length. Table~\ref{table:params} lists the various symbols used in this work. 
With no external disturbances, the thrust required by each quadcopter at hover is,
\begin{align}
	\droneThrust_i & = \| \droneMass_i g \mathbf{e_3} -  \tetherForce_{i} - \tetherForce_{i+1} \|, \label{eq:droneThrustTether}
\end{align}
where $m_i$ is the mass of the quadcopter and  $\tetherForce_{i}$, $\tetherForce_{i+1}$ are the catenary forces due to the tether on either side of the quadcopter.
\begin{table}[]
	\centering
	\begin{tabular}{|c | c |} 
		\hline
		Variables & Definition\\
		\hline\hline
		$\droneMass_j$ & Mass of Quadcopter $j$ $\msb{\SI{}{\kilogram}}$\\
		$\tetherLengthj$ & Tether length between Quadcopter $j{-}1$ and $j$ $\msb{\SI{}{\meter}}$ \\
		$\tetherMassPerLength_j$ & Tether mass per unit length of the $j$-th section $\msb{\SI{}{\kilogram \per \meter}}$\\
		$\tetherLength$ & Total tether length in the system $\msb{\SI{}{\meter}}$ \\
		$\droneThrust_j$ & Scalar thrust magnitude of Quadcopter $j$ $\msb{\SI{}{\newton}}$\\
		$\tetherForce\in \RealCoords$ & Force vector on the quadcopter due to the tether $\msb{\SI{}{\newton}}$\\
		\hline 
		$\sourcePower$ &Power supplied by the source $\msb{\SI{}{\watt}}$\\
		$\sourceVoltage$ & Voltage at the source $\msb{\SI{}{\volt}}$\\
		$\sourceCurrent$ & Current delivered by the source $\msb{\SI{}{\ampere}}$\\
		$\dronePowerj$ & Power consumption of Quadcopter $j$ $\msb{\SI{}{\watt}}$\\
		$\droneVoltagej$ & Voltage across Quadcopter $j$ $\msb{\SI{}{\volt}}$\\
		$\droneCurrentj$ & Current consumed by Quadcopter $j$ $\msb{\SI{}{\ampere}}$\\
		$\tetherResistancePerLengthj$ & Tether resistance per unit length of $j$-th section $\msb{\SI{}{\ohm \per \meter}}$\\
		$\tetherResistancej$ & Tether resistance of the $j$-th section $\msb{\SI{}{\ohm}}$ \\
		\hline
	\end{tabular}
	\caption{List of various symbols representing mechanical and electrical quantities used in this work.}
	\label{table:params}
\end{table}    

A uniform cable suspended between two fixed points forms a catenary curve\cite{lockwood1967book}, where the equation in the catenary plane is expressed using the catenary parameters $a,b,c$ as, 
\begin{align}
	z = a\cosh{\mrb{{(y-b)}/{a}}} +c, \label{eq:catenaryEquation}
\end{align}
and can be numerically solved using the position of the endpoints.
At hover, under quasi-static conditions, the tether between the quadcopters takes the catenary shape. The catenary parameter $a_i$ for each tether segment is computed by numerically solving the transcendental equation,
\begin{align}
	\sqrt{l_i^2-h_i^2} = 2a_i\sinh^2(\tfrac{d_i}{2a_i}), \label{eq:catTranscendentalEq}
\end{align}
and parameters $b_i, c_i$ are computed as follows,
\begin{gather}
	b_i = \tfrac{d_i}{2}{-}a_i \tanh^{-1}{(h_i/l_i)},
	c_i = {-} a_i\cosh{({-}b_i)/a_i}. 
\end{gather}

To solve for the catenary forces  $\tetherForce_{i}$, $\tetherForce_{i+1}$ at the ends of the tether segment, directions of the catenary forces at the ends are computed using the gradient of the catenary equation. 
The magnitude of the tensions is calculated by equating the net catenary forces to the weight of the tether (at quasi-static equilibrium).

Making use of the thrust generated by a quadcopter in \eqref{eq:droneThrustTether} at hover, we compute the electrical power consumption of the power train, using actuator disk model (see \cite{jain2020staging}) as, 
\begin{equation}
	\dronePoweri = \powerConst \droneThrust_i^{\frac{3}{2}},
	\label{eq:dronePower}
\end{equation}
where the power constant $\powerConst$ is a function of the propeller size $\propSize$, ambient air density $\airDensity$, propeller efficiency $\propEfficiency$ based on the propeller figure of merit, powertrain efficiency $\powertrainEfficiency$, and number of propellers $\nProps$,
\begin{equation}
	\powerConst = \frac{1}{\propEfficiency \powertrainEfficiency \sqrt{2 \airDensity \nProps \propSize}} .
\end{equation}
We assume that the propeller and powertrain efficiency is constant and neglects power consumption by the onboard sensors and computers.

To optimally design a power-supply system and to choose the various design factors for tethered quadcopters flight, we need to understand the effect of various design parameters which are considered in the next section.

\section{Electrical Power Analysis} 
\label{sec:systemAnalysis}

As mentioned in \secref{sec:intro}, power analysis can provide guarantees of being within rated limits of the power supply which ensures safety, and help in optimizing for design parameters such as tether resistance and input voltage.

\begin{figure}
	\centering
	\includegraphics[width=\columnwidth]{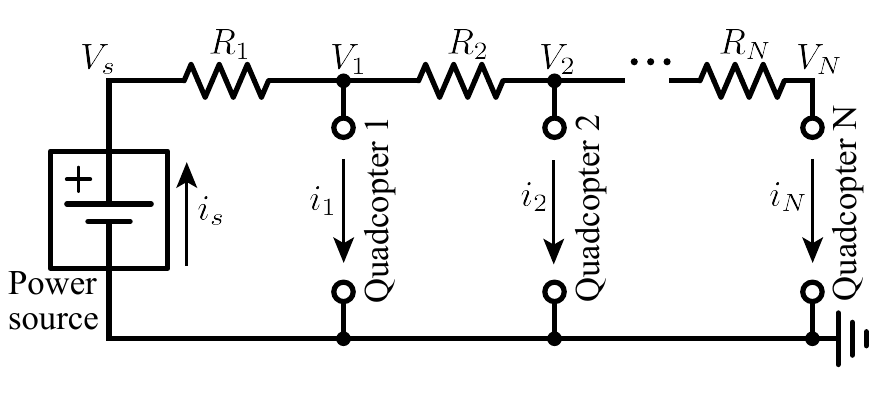}
	\caption{
		Circuit diagram for the proposed power supply architecture for $N$ tethered quadcopters.
		If the resistive losses are too high, a solution is to transmit power at a high voltage via the tether.
	}
	\label{fig:electricalSchematicMultiDrone}
\end{figure}

We analyze the power consumption and power supply requirements for $N$ quadcopters tethered to a single power supply with respect to the thrust that each quadcopter produces.
Consider $N$ quadcopters connected electrically in parallel as shown in \figref{fig:electricalSchematicMultiDrone}.
The portion of the tether between Quadcopter $(j-1)$ and Quadcopter $j$ has as a resistance $\tetherResistancej$.
The $j=1$ portion refers to the part of the tether between the power supply and Quadcopter~1.

The power supply voltage and current are denoted as $\sourceVoltage$ and $\sourceCurrent$ respectively.
We assume that the thrust to be produced by the $j$-th quadcopter $\droneThrust_j$ is already known.
Then from \eqref{eq:dronePower}, we know its power consumed $\dronePowerj$.
After the resistive voltage drop, let the voltage at the $j$-th quadcopter be $\droneVoltagej$.
Then the current consumption by this quadcopter will be given by,
\begin{equation}
	\droneVoltagej \droneCurrentj = \dronePowerj .
	\label{eq:dronePowerElectrical}
\end{equation}

Looking at \figref{fig:electricalSchematicMultiDrone}, we can apply Kirchoff's circuit laws to derive,
\begin{equation}
	\droneVoltagek = \droneVoltage_{k-1} - \mrb{\sum_{l=k}^{N} \droneCurrentl} \tetherResistancek, \quad k = 1, \ldots, N ,
	\label{eq:droneVoltageRecursive}
\end{equation}
where $\droneVoltage_{0} = \sourceVoltage$.
Simplifying the set of equations \eqref{eq:droneVoltageRecursive}, we can get the voltage across the quadcopters in terms of source voltage and current consumed by the quadcopters as,
\begin{equation}
	\droneVoltagej = \sourceVoltage - \sum_{k=1}^{j} \mrb{\sum_{l=k}^{N} \droneCurrentl} \tetherResistancek, \quad j = 1, \ldots, N .
	\label{eq:droneVoltageAbsolute}
\end{equation}

We have $2N$ unknowns $\mrb{\droneVoltage_1, \droneCurrent_1, \ldots, \droneVoltage_N, \droneCurrent_N}$ but only $N$ equations.
We know the power consumption $\dronePowerj$ by each quadcopter.
Therefore, we multiply each equation in \eqref{eq:droneVoltageAbsolute}, by the respective quadcopter current consumption $\droneCurrentj$, to get,
\begin{equation}
	\dronePowerj = \sourceVoltage \droneCurrentj - \droneCurrentj \sum_{k=1}^{j} \mrb{\sum_{l=k}^{N} \droneCurrentl} \tetherResistancek, \quad j = 1, \ldots, N .
	\label{eq:dronePowerSimultaneousQuadratics}
\end{equation}

Equations \eqref{eq:dronePowerSimultaneousQuadratics} are $N$ simultaneous quadratic equations in $N$ variables $\mcb{\droneCurrent_1, \ldots, \droneCurrent_N}$, which can be solved numerically.
With those values, we can get the power supplied $\sourcePower$ as,
\begin{equation}
	\sourceCurrent = \sum_{j=1}^N \droneCurrentj \;, \qquad  \sourcePower = \sourceVoltage \sourceCurrent .
\end{equation}

The analytical expression for the power supplied in terms of quadcopter thrust for $N=1$ is,
\begin{equation}
	\sourcePower = \frac{\sourceVoltage^2}{2 \tetherResistance_1} \mrb{1 - \sqrt{1 - \frac{4 \powerConst \droneThrust_1^{3/2} \tetherResistance_1}{\sourceVoltage^2}}}
	\label{eq:singleQuadPowerSupply}
\end{equation}
This expression is useful to draw conclusions about the dependence of thrust boundaries for the multiple quadcopter tethered system on the design parameters. 
This is covered in subsequent subsections.

\begin{figure}
	\centering
	\includegraphics[width=\columnwidth]{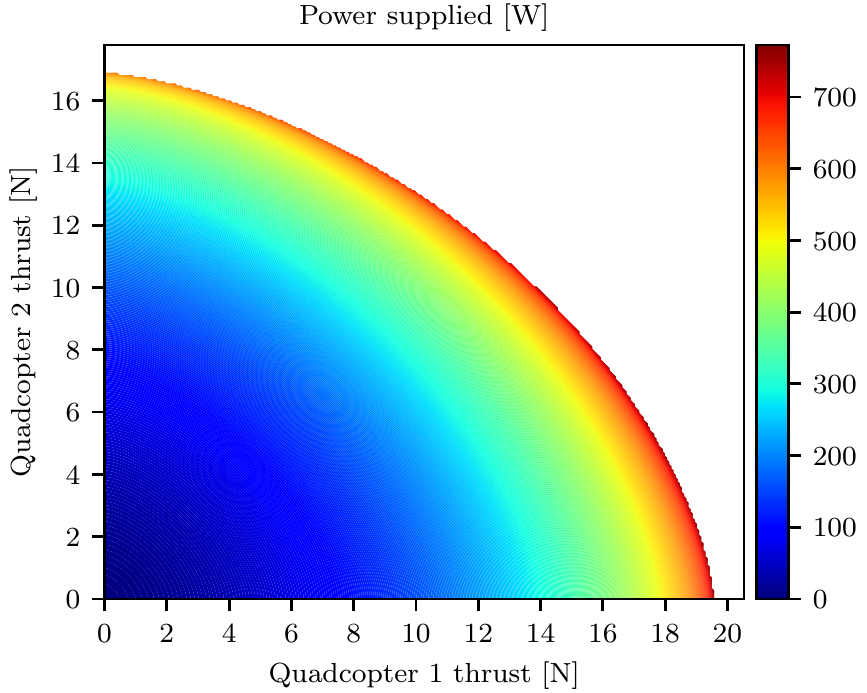}
	\caption{
		Source power $\sourcePower$ as a function of Quadcopter 1 thrust and Quadcopter 2 thrust.
		White regions indicate that the thrust combination is impossible to produce because of fundamental limitations.
		For this plot, the parameters are: $\sourceVoltage=$\SI{12.6}{\volt}, $12$ AWG wire $\mrb{\tetherResistancePerLength = \SI{0.0166}{\ohm \per \meter}}$, $l_1=\SI{6.1}{\meter}$, $l_2=\SI{1.5}{\meter}$, $\powerConst = \SI{4.5}{\watt \per \newton^{\frac{3}{2}}}$.
	}
	\label{fig:Ps_vs_quadThrusts}
\end{figure}

For the specific two quadcopters tethered system, the required supply power is plotted as a heatmap against the thrusts of the two quadcopters in \figref{fig:Ps_vs_quadThrusts}.
Once the system application is decided and the quadcopters are designed, such a plot could be used to choose the appropriate power supply and cable which meets the power requirements at the designed thrust values of the quadcopters.

Note that there is a boundary in the plot beyond which any additional current supplied would simply increase the resistive losses and none of the extra power will reach the quadcopters.
Thus any thrust combination that belongs in the white region in the plot cannot be produced.
We would like to emphasize that this is not a limitation of the power source, but a fundamental electrical limitation of the tethered-quadcopters system because of the physical parameters such as resistance and power consumption requirements.
This is also supported by equation \eqref{eq:singleQuadPowerSupply}, where increasing the thrust beyond a certain limit results in no real solutions for the supplied power.
This limit is given by,
\begin{equation}
	\droneThrust_\mathrm{crit} = \frac{\sourceVoltage^{4/3}}{\mrb{4 \powerConst}^{2/3} \tetherResistance ^{2/3}}
	\label{eq:thrustBoundary}
\end{equation}

Manipulating this thrust boundary could be of particular interest to designers to choose the power supply and tether size, given the system parameters, while guaranteeing electrical safety.
Given the quadcopters' power consumption coefficient $\powerConst$, the boundary can be influenced by the following parameters: power supply voltage $\sourceVoltage$ and the tether resistances $\mcb{\tetherResistance_1, \ldots, \tetherResistance_N}$.
This influence is studied in the following subsections.

\subsection{Feasible thrust boundary vs. source voltage}\label{sec:Ps_vs_Vs}
\begin{figure}
	\centering
	\includegraphics[width=.49\columnwidth]{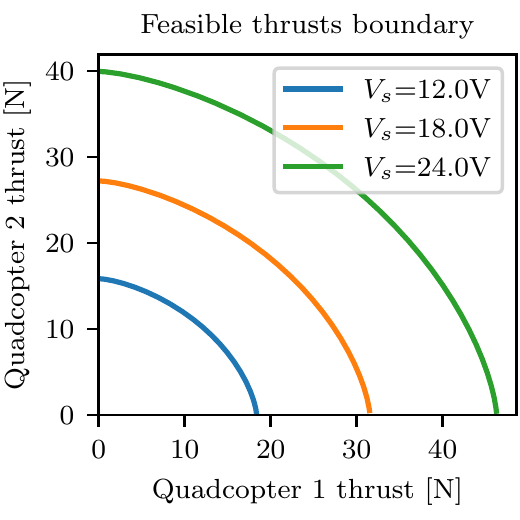}
	\includegraphics[width=.49\columnwidth]{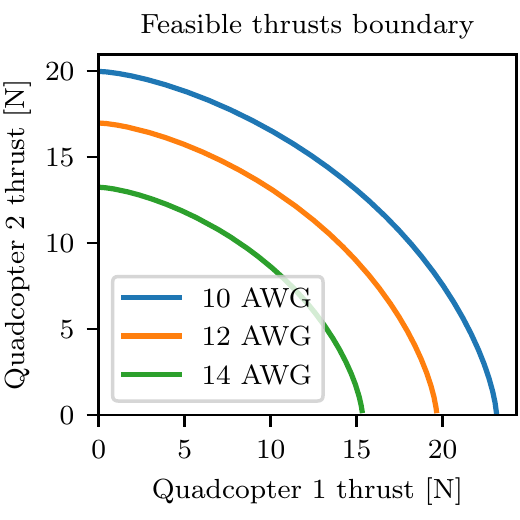}
	\caption{
		Maximum feasible thrust combinations for two quadcopters.\\
		(Left:) Different colored lines represent different source voltages $\sourceVoltage$.
		The boundary becomes larger as the source voltage is increased.
		Specifically, the dependence is $\sourceVoltage^{4/3}$.\\
		(Right:) Different colored lines represent different wire resistances.
		The boundary becomes larger as wire resistance per unit length $\tetherResistancePerLength$ is decreased.
		Specifically, the dependence is $\tetherResistancePerLength^{-2/3}$.
	}
	\label{fig:Ps_vs_Vs_AWG}
\end{figure}

We study the possible combinations of maximum thrusts that two tethered quadcopters can produce for various source voltages.
The plot of these boundaries can be seen in \figref{fig:Ps_vs_Vs_AWG} (Left).
The boundary is pushed outwards with increasing source voltage as $\sourceVoltage^{4/3}$, see \eqref{eq:thrustBoundary}, because the resistive losses are reduced on increasing the source voltage which allows for higher power consumption before reaching the fundamental limit.

\subsection{Feasible thrust boundary vs. wire size}\label{sec:Ps_vs_AWG}
The values of mass and resistance per unit length for different wire gauges are manufacturer dependent and are not explicitly known \textit{a priori}.
However, for a given manufacturer, we can measure the resistance and mass.
With decreasing wire gauge, the mass of the tether decreases, and the resistance increases.

For the analysis in this subsection, we assume the resistance per unit length $\tetherResistancePerLength$ of each section of the tether is the same.
We also assume that the length of each tether section is fixed to a particular value.
The plot of the feasible thrust boundaries for different wire gauges can be seen in \figref{fig:Ps_vs_Vs_AWG} (Right).
The boundary is pushed outwards with decreasing wire resistance per unit length as $\tetherResistancePerLength^{-2/3}$, see \eqref{eq:thrustBoundary}, because the resistive losses are reduced which allows for higher power consumption before reaching the fundamental limit.
There is no dependence on the mass per unit length of the tether because we are studying the behavior with respect to the thrust of the quadcopters.

\section{Horizontal Reach Power Comparison}\label{sec:catenary1v2}
\begin{table}
	\centering
	\begin{tabular}{|c|c||c|c|c|} 
		\hline
		& \thead{End effector \\ setpoint $[\SI{}{\meter}]$} & $\mrb{10,5}$ & $\mrb{20,10}$ & $\mrb{30,15}$ \\
		\cline{2-5}
		\multirow{3}{*}{\rotatebox[origin=c]{90}{\thead{One \\ quadcopter} \hspace{3.8mm}}} & $\sourceVoltage$ $[\SI{}{\volt}]$ & $18.0$ & $36.0$ & $54.0$ \\
		\cline{1-5}
		& \thead{Optimal tether \\ length $[\SI{}{\meter}]$} & $12.87$ & $26.08$ & $39.13$ \\
		\cline{2-5}
		& \thead{Minimum power $[\SI{}{\watt}]$} & \textbf{415.9} & \textbf{851.4} & \textbf{1459.6} \\
		\hline
		\multirow{3}{*}{\rotatebox[origin=c]{90}{\thead{Two \\ quadcopters} \hspace{1.3mm}}} & \thead{Optimal fraction} & 0.55 & 0.65 & 0.65 \\ \cline{2-5}
		& \thead{Optimal intermediate \\ setpoint $[\SI{}{\meter}]$} & \thead{$(6.26,$ \\ \hspace{1mm}$1.03)$} & \thead{$(14.34,$ \\ \hspace{1mm}$3.27)$} & \thead{$(21.21,$ \\ \hspace{1mm}$5.27)$} \\ \cline{2-5}
		& \thead{Minimum power $[\SI{}{\watt}]$} & \textbf{456.0} & \textbf{736.3} & \textbf{1083.0} \\
		\hline
	\end{tabular}
	\caption{
		Optimized power consumption comparison for one tethered quadcopter vs. two tethered quadcopter systems.
		The final quadcopter position, tether type $\mrb{\tetherResistancePerLength, \tetherMassPerLength}$, source voltage $\sourceVoltage$, total tether length $\tetherLength$ are common for the two cases.
	}
	\label{table:1v2Power}
\end{table}

In this section, we analyze the horizontal reach of the tethered quadcopter system for two cases: (i) One tethered quadcopter, and (ii) Two tethered quadcopters.
The performance metric is the power that needs to be supplied for the system at hover, with the final (end effector) quadcopter in the tether series being at a specific desired position.

Catenary forces and the quadcopter thrust are computed as described in \secref{sec:description}.
Although adding a second quadcopter to a single tethered quadcopter system requires supporting its additional mass, the weight and tension of the tether getting distributed between the two quadcopters could result in a reduction of total power -- this is because the power consumption $\dronePower$ is dependent on the $1.5$th power of the thrust $\droneThrust^{3/2}$.

The approach to compare the total power consumption for the two cases is as follows:
\begin{enumerate}
	\item Fix the tether type (mass per unit length $\tetherMassPerLength$ and resistance per unit length $\tetherResistancePerLength$) and the source voltage and assume all quadcopters on the tether are identical.
	\item Choose a desired end effector setpoint -- the horizontal and vertical distance from the tether anchor which is at the power supply. Note the end effector setpoint for the one quadcopter case is given as $(d_1, h_1)$ and for two quadcopters it is $(d_1{+}d_2, h_1{+}h_2)$, where $d_1, h_1, d_2, h_2$ are as shown in \figref{fig:dynamicsSchematicMultiDrone}.
	\item Consider the one tethered quadcopter case and choose an optimal tether length that minimizes the power to be supplied.
	\item With the computed optimal length as the total tether length, add a second (intermediate) quadcopter and minimize power with respect to the following parameters: (i) intermediate quadcopter setpoint (from the power supply), (ii) fraction of the tether length between the power supply and intermediate quadcopter.
	\item Compare the optimized power consumption values for the two cases.
\end{enumerate}

Note that the power supply, the intermediate quadcopter, and the end effector quadcopter must all be in the same vertical plane, since any deviations of the intermediate quadcopter perpendicular to the vertical plane will increase the tension in that direction, leading to a higher thrust and power consumption for the intermediate quadcopter.
Therefore, this is a 2D problem.

In all the problems in this analysis, we assume that the quadcopter mass is \SI{0.7}{\kilogram}, and the tether we use is $12$ AWG for which $\tetherMassPerLength=\SI{0.095}{\kilogram \per \meter}$ and $\tetherResistancePerLength=\SI{0.0166}{\ohm \per \meter}$.
The power constant is experimentally determined to be $\powerConst = \SI{4.5}{\watt \per \newton^{\frac{3}{2}}}$.

\begin{figure}
	\centering
	\includegraphics[width=\columnwidth]{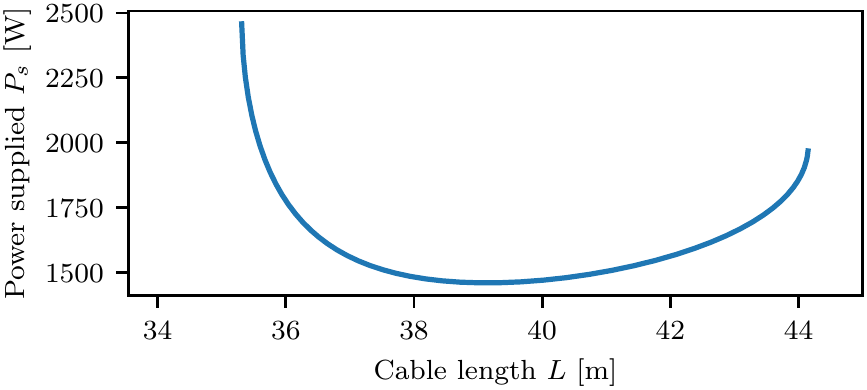}
	\caption{
		A plot of required power to be supplied to a single tethered quadcopter to hover at $\mrb{30,15}\SI{}{\meter}$ vs. tether length.
	}
	\label{fig:one_drone_power_vs_tether_length}
\end{figure}

\begin{figure}
	\centering
	\includegraphics[width=\columnwidth]{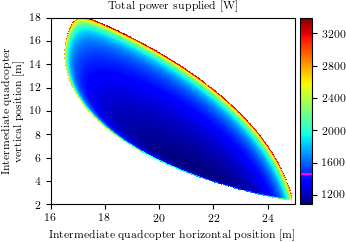}
	\caption{
		A plot of total power to be supplied to a two quadcopter tethered system vs. intermediate quadcopter horizontal and vertical setpoint.
		The end effector quadcopter is taken to be at $\mrb{30,15}\SI{}{\meter}$.
		The plot is for the optimized fraction of tether length.
		Pink colored line on the colorbar shows optimized power consumption for the one quadcopter system for the same end effector setpoint.
	}
	\label{fig:sweep_multidrone_ix_iz_frac}
\end{figure}

\begin{figure}
	\centering
	\includegraphics[width=\columnwidth]{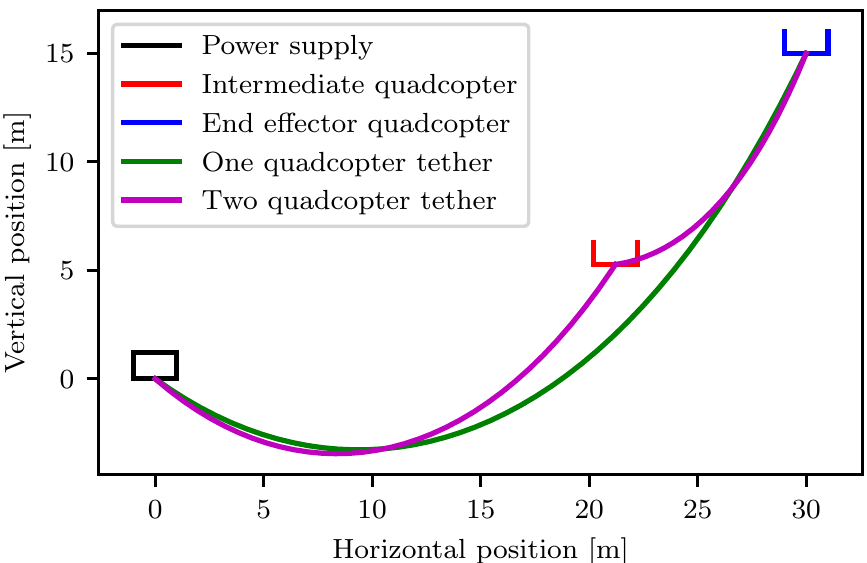}
	\caption{
		Sketch comparing one tethered quadcopter and two tethered quadcopters for end effector setpoint of $\mrb{30,15}\SI{}{\meter}$.
		The two quadcopter system consumes lesser total power than one quadcopter.
	}
	\label{fig:one_drone_vs_two_drone_sketch}
\end{figure}

\begin{figure*}
	\centering
	\begin{subfigure}[b]{\textwidth}
		\centering
		\includegraphics[width=\textwidth]{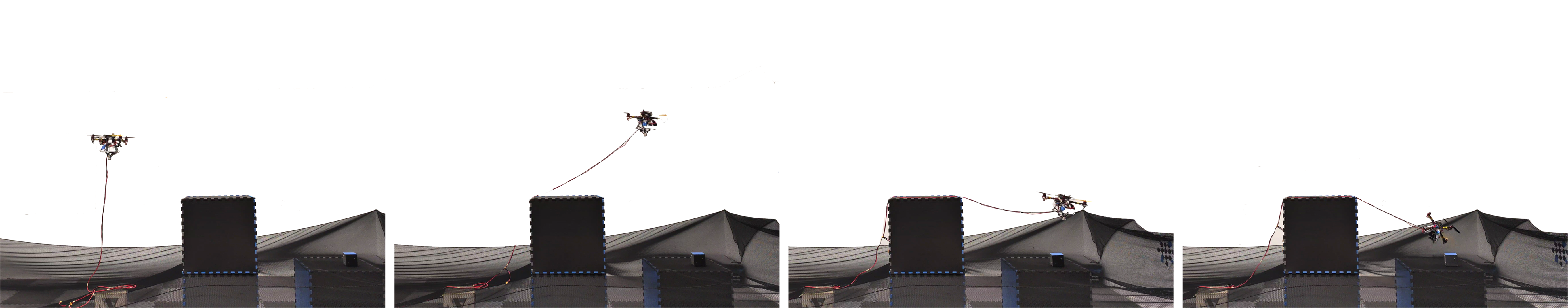}
		\caption{Snapshots of a grasping attempt using a single tethered quadrotor (setup {\it(i)}) over an obstacle -- this ends up crashing the quadcopter}
		\label{fig:singleTetherGrasp}
	\end{subfigure}
	
	\begin{subfigure}[b]{\textwidth}
		\centering
		\includegraphics[width=\textwidth]{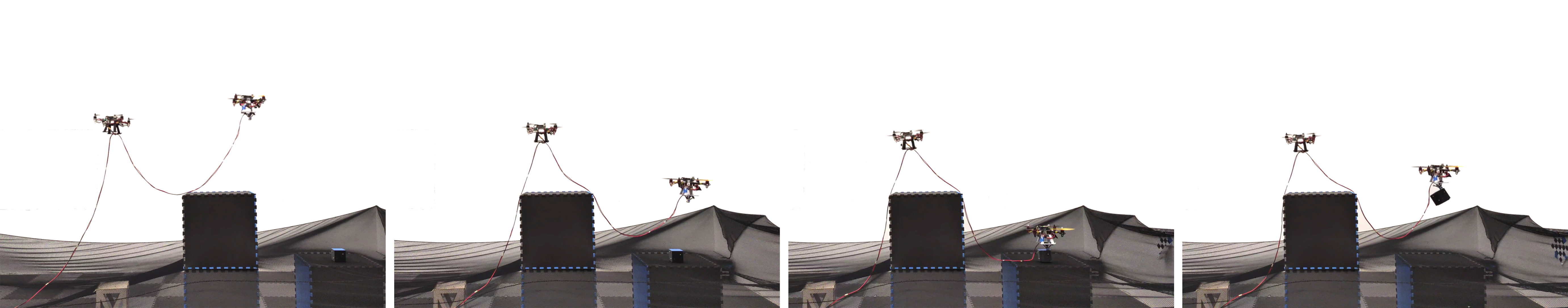}
		\caption{Snapshots of two quadcopters connected by single tether (setup {\it(ii)}) grasping over an obstacle}
		\label{fig:dualTetherGrasp}
	\end{subfigure}
	\caption{Experimental demonstrations of tethered drones for grasping over obstacles}
	\label{fig:grasping_example}
\end{figure*}

Results for three end effector setpoints -- $\mcb{\mrb{10,5},\mrb{20,10},\mrb{30,15}}\SI{}{\meter}$ are summarized in \tabref{table:1v2Power}.
The setpoints are chosen to cover a wide range of tether lengths.
In the first case, the horizontal reach and tether weight is relatively low, making the one quadcopter tether system more power efficient than the two quadcopter tether system.
In the second case, we require a higher horizontal reach increasing the catenary force on the end effector quadcopter.
Adding an intermediate quadcopter reduces the total power consumption slightly (by $13.5\%$).

For the third case, the system requires a long horizontal reach and needs to lift a heavy tether ($\sim \SI{4}{\kilogram}$).
Adding the intermediate quadcopter results in the load distributed among the two quadcopters.
This results in $26\%$ lesser total power consumption for the two quadcopter system as compared to the one quadcopter system.
Adding an intermediate quadcopter for such high tether lengths also frees up some thrust capacity for the end effector quadcopter, making it more maneuverable and agile.
In applications such as firefighting, higher agility could be very useful in controlling the water outlet precisely and moving it rapidly.

A visual comparison of the one quadcopter and two quadcopter systems along with the tether shapes for the case with end effector setpoint $\mrb{30,15}\SI{}{\meter}$ is shown in \figref{fig:one_drone_vs_two_drone_sketch}.

For end effector setpoint of $\mrb{30,15}\SI{}{\meter}$, the plot of power consumption of one quadcopter tethered system with respect to the tether length is shown in \figref{fig:one_drone_power_vs_tether_length}.
We take the point with the minimum power consumption for step (3) in the approach.
Note that a minimum always exists because for low tether lengths, the tension due to catenary force is high, and at high tether lengths, the weight of the tether is high -- both resulting in higher power consumption.
The plot of total power consumption of the two quadcopter tethered system with respect to the intermediate quadcopter setpoint (for the optimized fraction of tether length) is shown in \figref{fig:sweep_multidrone_ix_iz_frac}.
Using a similar argument as for the one quadcopter system, we claim that there exists a minimum value of total power consumption at some optimal value of the intermediate quadcopter setpoint and the fraction of tether length between the power supply and intermediate quadcopter.
\section{Experimental Demonstrations}
In this section, we present experiments demonstrating potential applications for two quadcopters powered using a single tether. A two quadcopter system has the advantages of a single tethered quadcopter, such as extended flight time, and a steady power supply, while also increasing maneuverability in the horizontal direction, especially over difficult/unknown terrains. In the rest of the section, we present two experiments, comparing the following setups,
\begin{enumerate}[label=\emph{(\roman*)}]
	\item a single tethered quadcopter.
	\item a two quadcopter system powered by a single tether
\end{enumerate}

\subsection{Passing Through Windows}

In this experiment, we consider a window passing example with tethered quadcopters as shown in \figref{fig:introduction}. In setup {\emph{(ii)}, the first quadcopter (left)  supports the second quadcopter (right) exploring on the other side of the window. Similar exploration using single tethered quadcopter are not always feasible in cluttered environments, as the obstacles could interfere with the tether.

\subsection{Grasping Over Obstacles} 

A two quadcopter tethered system can act as an \emph{aerial series-manipulator} in cluttered environments, as shown in \figref{fig:dualTetherGrasp}. We do a grasping experiment using the two setups, to grasp an object located on the other side of an obstacle, in the vertical plane as shown in \figref{fig:grasping_example}. Waypoints are provided to the quadcopters to reach over the obstacles and grasp the object.
An electromagnet is used as the gripper for grasping the metallic object.
In setup {\emph{(i)}, see \figref{fig:singleTetherGrasp}, a single tethered quadcopter attempts to grasp the object, however, is unable to reach the grasp location due to the limitation of the quadcopter to drag the tether over obstacles. 
In setup {\emph{(ii)}, the first quadcopter (left) acts an intermediate joint for the second drone (right) to help grasp the object over the obstacle as shown in \figref{fig:dualTetherGrasp}. 
A video demonstrating the experiments is provided in the media attachment.

\section{Conclusion and Future Work}\label{sec:conclusion}
In this paper, we have presented an analysis for a tethered multiple-quadcopter system to estimate the electrical power consumption by the quadcopters and the power supply required to deliver the desired power to the quadcopters.
The analysis includes various mechanical, aerodynamic, and electrical parameters of the tethered multiple-quadcopter system.
We analyzed the power supply requirement for multiple tethered quadcopters with respect to the quadcopter thrusts.
We found that there exists a thrust boundary that cannot be exceeded because of fundamental electrical limitations.
We found that the thrust boundary can be pushed outward by increasing the supply voltage or decreasing the tether resistance which can be helpful to designers.

We compared the power requirements for various end effector setpoints for one tethered quadcopter and a series of two tethered quadcopters.
As the end effector setpoint is pushed further from the power supply (the anchor point), it was found that the optimal two quadcopter system consumes lesser total power than the optimal one quadcopter system.
Adding an intermediate quadcopter also frees up thrust capacity on the end effector quadcopter increasing its maneuverability and agility.

Lastly, we show a system with two experiments using two quadcopter tethered system and demonstrate better horizontal reachability in cluttered environments as compared to the corresponding one quadcopter system.
An additional advantage offered by this system is performing tasks that need quadcopters to operate in proximity such as picking fruits or cleaning a high-rise building.

An extension to this work is to design a multiple quadcopter tether system to perform manipulation tasks in cluttered environments. Using the power analysis allows users to specify the power supply and tether sizes optimally for the specific system.

\section*{Acknowledgment}
The experimental testbed at the HiPeRLab is the result of the contributions of many people, a full list of which can be found at \url{hiperlab.berkeley.edu/members/}.

\newpage
\balance
\bibliographystyle{IEEEtran}
\bibliography{root.bbl}

\end{document}